\documentclass[10pt, conference]{IEEEtran} 

\usepackage{url}
\usepackage{makecell}
\usepackage{amsfonts}
\usepackage{amssymb}
\usepackage{amsmath}
\usepackage{graphicx, colordvi, psfrag}
\usepackage{calc,pstricks, pgf, xcolor}
\usepackage{epsfig, cite}
\usepackage{bm} 
\usepackage{bbm} 
\usepackage{enumerate}

\newcommand{\beq}[1]{\begin{equation}\label{#1}}
\newcommand{\eeq}{\end{equation}}

\newcommand{\beqn}[1]{\begin{eqnarray}\label{#1}}
\newcommand{\eeqn}{\end{eqnarray}}

\newtheorem{thmbody}{Theorem}
\newenvironment{thm}{
\begin{thmbody}
	}{
	\end{thmbody} 
	}
\newtheorem{dfnbody}{Definition}

\newtheorem{corbody}{Corollary}

\newtheorem{lemmabody}{Lemma}
\newenvironment{lemma}{
\begin{lemmabody}
	}{
	\end{lemmabody} 
	}
\newtheorem{propbody}{Proposition}

\newenvironment{proof}{
	{\it Proof:}
	}{
 $\Box$
	}

\usepackage{dsfont}
\begin{document}
\title{Channel Input Adaptation\\ via\\ Natural Type Selection}


\author{\IEEEauthorblockN{Sergey Tridenski}
\IEEEauthorblockA{EE - Systems Department\\Tel Aviv University\\
Tel Aviv, Israel\\
Email: sergeytr@post.tau.ac.il}
\and
\IEEEauthorblockN{Ram Zamir}
\IEEEauthorblockA{EE - Systems Department\\Tel Aviv University\\
Tel Aviv, Israel\\
Email: zamir@eng.tau.ac.il}}



\maketitle
\begin{abstract}
For the model of communication through a discrete memoryless channel using i.i.d. random block codes,
where the channel is changing slowly from block to block,
we propose a stochastic algorithm for adaptation of the generating distribution of the code in the process of continuous reliable communication.
The purpose of the algorithm is to match the generating distribution $Q(x)$ to the changing channel $P(y\,|\,x)$,
so that reliable communication is maintained at some constant rate $R$.
This is achieved by a feedback of one bit per transmitted block.
The feedback bit is determined by the joint type of the last transmitted codeword and the received block,
a constant threshold $T>R$, and some conditional distribution $\Phi(x\,|\,y)$.
Depending on the value of the feedback bit, the system parameters $Q(x)$ and $\Phi(x\,|\,y)$ are both updated according to the joint type
of the last transmitted and received blocks, or remain unchanged.

We show that, under certain technical conditions,
the iterations of the algorithm lead to a distribution $Q(x)$, which guarantees reliable communication for all rates below the threshold $T$,
provided that the discrete memoryless channel capacity of $P(y\,|\,x)$ stays above $T$.
\end{abstract}




%
%
%
%



\section{Introduction} \label{Intro}
\bigskip

Consider a standard information theoretic scenario of communication through a discrete memoryless channel $P(y\,|\,x)$ using block codes.
For this case, information theory provides optimal solutions in the form of the channel input distribution ${Q\mathstrut}^{*}(x)$, achieving the Shannon capacity $C$,
or achieving the Gallager error exponent $E(R)$ for a given communication rate $R$.
Suppose, however, that the channel stochastic matrix $P(y\,|\,x)$ is slowly, or rarely, changing with time and we would like to sustain reliable communication at a constant rate $R$. For this purpose, we assume
using a single bit of feedback, from the receiver to the transmitter, per transmitted block.
In our model, we further assume that,
given this bit of feedback,
the system parameters are updated
using the last transmitted (and received) block only,
i.e. without memory from the previous blocks.
So that, potentially, the system will follow the changes in the channel more closely.
Our goal of sustaining reliable communication at a constant rate R is legitimate and feasible, of course, only as long as the capacity of the channel $C$, as a function of $P(y\,|\,x)$, stays above the rate $R$.
While the channel capacity may stay well above the rate, the optimal solution ${Q\mathstrut}^{*}(x)$ may drift significantly, as a result of the drift in $P(y\,|\,x)$, and render the initial code unreliable.

In this work, the block code is modeled as a random code, generated i.i.d. with a distribution $Q$.
The reason for modeling the code as an i.i.d. random code is twofold.
First, random codes
achieve capacity.
The idea is to fix some intermediate $T>R$ and, by changing $Q$, to keep the {\em correct-decoding} random coding exponent
\cite{Arimoto76}, \cite[eq.~31]{TridenskiZamir17}, determined by $Q$, ``pinned'' to zero at a rate $R'=T$,
provided that $T<C$.
This would mean that the corresponding {\em error} exponent ${E}_{r}(R', Q)$
\cite[eq.~(5.6.28)]{Gallager} is strictly positive for all $R'<T$, thus
ensuring, in particular, reliable communication at $R$.

Secondly, an i.i.d. distribution in a random code,
as opposed, for example, to a uniform distribution over a single type, results in a certain diversity of the codeword types,
which allows us to invoke a mechanism of {\em natural type selection} for update of the parameter $Q$.
Using this mechanism iteratively, we successively update the codebook distribution $Q$ so that eventually the correct-decoding exponent, associated with $Q$, decreases to zero at $T$, thus achieving our goal.

The mechanism of natural type selection (NTS) has been originally observed and studied in the lossy source coding setting \cite{ZamirRose01}, \cite{KochmanZamir02}.
In that setting, a discrete memoryless source is mapped into a reproduction codebook, generated i.i.d. according to a distribution $Q$.
In the encoding process,
a linear 
search is performed through the codebook, until the first reproduction sequence is found, which
is close enough
to the source sequence.
Since various types are inherently present in the i.i.d. codebook,
the empirical distribution of the winning reproduction sequence, in general, is different than $Q$, and
is used for generation of the next codebook.
This results in a decrease in the compression rate, which, after repeated iterations, converges to the optimum, given by the rate-distortion function.
This last property is guaranteed by the fact that both the conditional type, given the source sequence,
and the marginal type, of the winning sequence, with high probability,
evolve along two analogous steps of the Blahut algorithm for rate-distortion function computation \cite{Blahut72}.

In our previous attempt to find a parallel NTS phenomenon in channel coding \cite{TridenskiZamir15}
we looked for a stochastic counterpart of the Arimoto algorithm for random coding exponent computation \cite{Arimoto76}.
One of a number of difficulties there
remains
setting 
a slope $\rho$ of the exponent,
which is a constant parameter in the Arimoto algorithm.
A viable alternative to this could be some kind of a ``variable slope'' version of the Arimoto algorithm,
as in the case of the Blahut algorithm for source coding.

In the current paper, we abandon the exact steps of the Arimoto algorithm,
retaining nominally its two components -- the channel input distribution, denoted here as $Q(x)$, and a conditional distribution $\Phi(x\,|\,y)$.
In the stochastic algorithm, proposed here, $Q(x)$ becomes updated by the type of a ``good'' transmitted sequence $\bf x$, and $\Phi(x\,|\,y)$
is updated by the conditional type of $\bf x$, given the corresponding received sequence $\bf y$.
The ``goodness'' of the transmitted sequence is determined at the decoder
by its joint type with the received sequence,
a threshold $T$, and using $\Phi(x\,|\,y)$.
The channel $P(y\,|\,x)$ itself is changing slowly/rarely and is assumed constant during iterations.

The details of the proposed scheme are given in Section~\ref{Scheme}.
Section~\ref{Type} serves as a bridge between the stochastic procedure and the underlying non-stochastic algorithm.
Section~\ref{Iterations} contains our main result, stating convergence of the iterations.
In Section~\ref{Discussion} we discuss the convergence result and assumptions we have to make.

\bigskip

\section{Adaptation scheme} \label{Scheme}
\bigskip

Let $P(y\,|\,x)$ be a discrete memoryless channel with finite input and output alphabets $\cal X$ and $\cal Y$, respectively, and suppose we use a random codebook of blocklength $n$ and size $e^{nR}$, generated i.i.d. according to a distribution $Q(x)$,
for communication through this channel. We assume that the rate $R$ is sufficiently lower than the mutual information $I(X; Y) \,\equiv\,I\big(Q(x)\cdot P(y\,|\,x)\big)\,\equiv\,I(Q\circ P)$,
so that the decoding error exponent\footnote{Expressed here in a common framework with our results \cite[eq.~28]{TridenskiZamir17}. Throughout the paper, we use notations $U(x, y)$ and $U_{x,\,y}$ interchangeably,
also for the marginal and conditional distributions, e.g. $U(x)$ and $U_{x}$. The notation $D(\cdot\|\cdot)$ stands for the information divergence.} \cite{Gallager}:
\begin{align}
{E}_{r}(R, Q) & = \min_{\substack{\\U(x, \, y)}} \Big\{D({U\!\mathstrut}_{x, \, y}\, \| \, Q\circ P)
\nonumber \\
& \;\;\;\;\;\;\;\;\;\;\;\;\;\;\;\;
+\, \big|
D({U\!\mathstrut}_{x, \, y}\,\|\,Q \times {U\!\mathstrut}_{y})
- R\big|^{+}\Big\}
\label{eqErrorExponent}
\end{align}
is sufficiently high for our purposes.

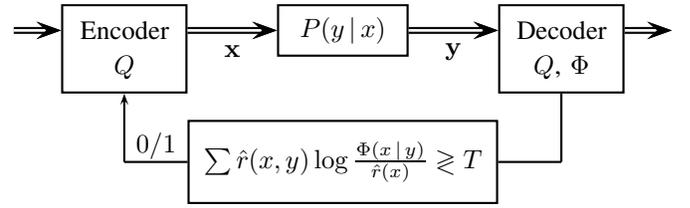
\begin{figure}[!htb]
\psset{unit=.7mm}
\begin{center}
\begin{pspicture}(0, 32)(125, 70)

\psframe(9, 53)(33, 70)   \rput(21, 65){Encoder}\rput(21, 58){$Q$}
\psline[doubleline = true]{->}(0, 65)(9, 65)
\psline[doubleline = true]{->}(33, 65)(50, 65)

\rput(41.5, 61){${\bf x}$}

\psframe(50, 60)(75, 70) \rput(62.5, 65){$P(y\,|\,x)$}

\rput(83.5, 61){${\bf y}$}

\psline[doubleline = true]{->}(116, 65)(125, 65)
\psline[doubleline = true]{->}(75, 65)(92, 65)
\psframe(92, 53)(116, 70) \rput(104, 65){Decoder}\rput(104, 58){$Q$, $\Phi$}

\psframe(33, 32)(92, 48)
\rput(62.5, 40){$\sum \hat{r}(x, y)\log\frac{\Phi(x\,|\,y)}{\hat{r}(x)}\gtrless T$}

\psline{-}(104, 53)(104, 40)
\psline{-}(92, 40)(104, 40)

\psline{-}(21, 40)(33, 40)
\psline{<-}(21, 53)(21, 40)

\rput(27, 43){$0 / 1$}

\end{pspicture}
\end{center}
\caption{Channel with a $1$-bit feedback.}
\label{fig1}
\end{figure}

With high probability, the decoder guesses the sent codeword correctly.
Let $r(x, y)$ denote the joint type of the transmitted and the received blocks of length $n$, both of which are available at the decoder after correct decoding,
so that the estimated joint type at the decoder is $\hat{r}(x, y) \, = \, r(x, y)$.
The decoder then sends reliably a bit of feedback, $F \, = \, 0$ or $1$, to the transmitter, according to the following rule:
\begin{equation} \label{eqAckNAk}
\sum_{x, \, y} \hat{r}(x, y) \log \frac{\Phi(x \, | \, y)}{\hat{r}(x)} \;\; \geq \;\; T
\;\;\;\;\;\;
\Longleftrightarrow
\;\;\;\;\;\;
F \, = \, 1,
\end{equation}
where $\hat{r}(x)\,=\,\sum_{y}\hat{r}(x, y)$ represents the estimated type of the sent codeword, $\Phi(x \, | \, y)$ is some fixed conditional distribution, 
known at the decoder side, and $T$ is a real number (Fig.~\ref{fig1}).

In case $F = 1$, which is a rare event for a large enough threshold $T$, the system parameters are updated: a new codebook
is adopted by both the encoder and the decoder, generated according to a new distribution $\,Q'(x) = r(x) = \hat{r}(x)\,$,
known at both sides,
and the decoder chooses the conditional type $\hat{r}(x \, | \, y)$ as its new stochastic matrix $\Phi'(x \, | \, y)$.
If the type of the received block is $\,\hat{r}(y) = 0$ on some letter $y\,\in\,{\cal Y}$, then the corresponding conditional distribution given this letter, $\Phi(x \, | \, y)$,
remains
unchanged.
In case the feedback $F = 0$, both system parameters $Q$ and $\Phi$ remain unchanged.
To summarize:
\begin{displaymath}
\begin{array}{c|c|c}
\text{Feedback} & \text{Encoder} & \text{Decoder} \\
& & \\[-1em]
\hline
& & \\[-1em]
F\mathstrut\,=\,1 & Q(x)\,\leftarrow \, r(x) & Q(x)\,\leftarrow \, \hat{r}(x) \\
&
\;\;\;\;\;\;\;\;\;\;\;\;\;\;\;\;\;\;\;\;\;\;\;\;\;\;\;\;\;\;
&
\Phi(x\,|\,y) \,\leftarrow\,\hat{r}(x\,|\,y) \\
& & \\[-1em]
\hline
& & \\[-1em]
F\,=\,0 & - & -
\end{array}
\end{displaymath}


\bigskip

\section{Convergence of a type} \label{Type}
\bigskip

The joint type $\hat{r}(x, y)$ is related, of course, to the exponent in the probability of the event $\{F = 1\}$,
the {\em update} exponent:

\bigskip
{\em Proposition 1:}
{\em Given the event $\{F = 1\}$, as the blocklength $n$ increases,
with high probability $\hat{r}(x, y) = r(x, y)$, and this type
converges in probability to a distribution ${U\mathstrut}^{*}(x, y)$, which is the unique solution of the minimum}
\begin{equation} \label{eqErasureExponent}
\hat{E}(T, Q, \Phi) \,\triangleq \min_{\substack{\\U(x, \, y):\\\sum_{x, \, y}U(x, \,y) \log\frac{\Phi(x \, | \, y)}{U(x)} \; \geq \; T}}
\!\!
D({U\!\mathstrut}_{x, \, y} \, \| \, Q\circ P).
\end{equation}
{\em Provided that
the error exponent of the decoder\footnote{The decoding error exponent may be lower than (\ref{eqErrorExponent}), depending on the specific decoder/sequence of decoders as a function of $n$, which we use.}, e.g. (\ref{eqErrorExponent}), is higher than (\ref{eqErasureExponent}).}

\bigskip

\begin{proof}
Observe that (\ref{eqErasureExponent}) is in fact the exponent of the event, pertaining to the {\em true} type $r(x, y)$, regardless of the decoding success:
\begin{displaymath}
\bigg\{\sum_{x, \, y} r(x, y) \log \frac{\Phi(x \, | \, y)}{r(x)} \;\; \geq \;\; T \bigg\}.
\end{displaymath}
If the decoding error exponent is higher than (\ref{eqErasureExponent}),
then the analogous event for $\hat{r}(x, y)$ has also the exponent given by (\ref{eqErasureExponent}). The convergence of the type in probability can be shown using Sanov's theorem \cite{Cover}.
\end{proof}


Although unnecessary here,
it can be checked that the update exponent (\ref{eqErasureExponent})
is an upper bound on the correct-decoding exponent,
associated with $Q$, \cite[eq.~32]{TridenskiZamir17} at $R'=T$.
Moreover, (\ref{eqErasureExponent}) coincides with the optimal correct-decoding exponent for the optimal $Q$ and $\Phi$
of the Arimoto algorithm \cite{Arimoto76}.

In what follows, we disregard completely the stochastic nature of the type $\hat{r}(x, y)$
(i.e. assume that the blocklength $n$ is large enough)
and the possibility that
the update exponent (\ref{eqErasureExponent}) exceeds the decoding error exponent, e.g. (\ref{eqErrorExponent}).
We assume simply
that $\hat{r}(x, y) = r(x, y) = {U\mathstrut}^{*}(x, y)$,
and examine convergence properties of the sequence of iterative solutions ${U\mathstrut}^{*}_{l}(x, y)$, $l = 0$, $1$, $2$, ... , of (\ref{eqErasureExponent}).

\bigskip

\section{Convergence of iterations} \label{Iterations}
\bigskip

\begin{lemma} \label{lemma1}
{\em Let the exponent (\ref{eqErasureExponent}) be finite for some $T$, $Q = {Q\mathstrut}_{0}$ and $\Phi = {\Phi\mathstrut}_{0}\,$. An iterative update of the parameters $Q$ and $\Phi$ in (\ref{eqErasureExponent}) by the corresponding solution ${U\mathstrut}^{*}(x, y)$:}
\begin{align}
{Q\mathstrut}_{l\,+\,1}(x) \; & \leftarrow \; {U\mathstrut}^{*}_{l}(x),
\label{eqUpdateQ} \\
{\Phi\mathstrut}_{l\,+\,1}(x \, | \, y) \; & \leftarrow \;
\left\{
\begin{array}{l l}
{U\mathstrut}^{*}_{l}(x \, | \, y), & \text{if} \;\;\; {U\mathstrut}^{*}_{l}(y) \, > \, 0 \\
{\Phi\mathstrut}_{l}(x \, | \, y), & \text{if} \;\;\; {U\mathstrut}^{*}_{l}(y) \, = \, 0
\end{array}
\right.
\label{eqUpdatePhi}
\end{align}
{\em results in a monotonically non-increasing sequence $\big\{\hat{E}(T, {Q\mathstrut}_{l}, {\Phi\mathstrut}_{l})\big\}_{l\, = \, 0}^{+\infty}\;$ of (\ref{eqErasureExponent})}.
\end{lemma}

\bigskip

\begin{proof}
Observe that the divergence in (\ref{eqErasureExponent}) can be broken up into two terms: the exponent of the codeword type -- $D\big(U(x) \, \| \, Q(x)\big)$, and the conditional exponent, given the distribution $U(x)$. The second term can be minimized over $U(y \, | \, x)$ separately for each distribution $U(x)$, and the interesting property is that the resulting conditional exponent given $U(x)$ has no dependence on $Q$. Therefore, we can reduce the first term $D\big({U\mathstrut}^{*}(x) \, \| \, Q(x)\big)$ in the minimum independently (to zero) by replacing $Q(x)$ with ${U\mathstrut}^{*}(x)$. The second term in the minimum, which is the conditional exponent given ${U\mathstrut}^{*}(x)$, will stay the same given the same ${U\mathstrut}^{*}(x)$, and can only be reduced further (given ${U\mathstrut}^{*}(x)$) by replacing $\Phi(x\,|\,y)$ in the minimization condition with ${U\mathstrut}^{*}(x \, | \, y)$, simply because the previous achieving joint distribution ${U\mathstrut}^{*}(x, y)$ will satisfy the new condition as well.
Minimizing both terms over $U(x)$ again, we further reduce the result. Formally:
\begin{align}
& \hat{E}(T, Q, \Phi)
\nonumber \\
& = \min_{\substack{\\U(x, \, y):\\\sum_{x, \, y}U(x, \, y) \log\frac{\Phi(x \, | \, y)}{U(x)} \; \geq \; T}}
\!\!\!\!\!
\Big\{D({U\!\mathstrut}_{x} \, \| \, Q) +
D({U\!\mathstrut}_{x,\,y} \,\|\, {U\!\mathstrut}_{x}\circ P)
\Big\}
\nonumber \\
& = \;\;\;\;\;\;\;\;\;\;\;\;\;\;\;\;\;\;\;\;\;\;\;\;\;\;\;\;\;\;\;\;\;\;
\;\;
\underbrace{D\big({U\mathstrut}^{*}_{x} \, \| \, Q\big)}_{\geq \, 0} +\,
D({U\mathstrut}^{*}_{x,\,y}\,\|\, {U\mathstrut}^{*}_{x}\circ P)
\nonumber \\
& \!\overset{(a)}{\geq}
\;\;\,
\min_{\substack{\\U(y \, | \, x):\\ \sum_{x, \, y}{U\mathstrut}^{*}(x)U(y \, | \, x) \log\frac{{U\mathstrut}^{*}(x \, | \, y)}{{U\mathstrut}^{*}(x)} \\ \geq \\ \sum_{x, \, y}{U\mathstrut}^{*}(x, \, y) \log\frac{{U\mathstrut}^{*}(x \, | \, y)}{{U\mathstrut}^{*}(x)}}}
\;\;\,
D\big({U\mathstrut}_{x}^{*} \circ {U\!\mathstrut}_{y \, | \, x}\,\|\,{U\mathstrut}_{x}^{*}\circ P\big)
\nonumber \\
& =
\;\;\;
\min_{\substack{\\U(y \, | \, x):\\ \sum_{x, \, y}{U\mathstrut}^{*}(x)U(y \, | \, x) \log\frac{{U\mathstrut}^{*}(x \, | \, y)}{{U\mathstrut}^{*}(x)} \\ \geq \\ \sum_{x, \, y}{U\mathstrut}^{*}(x, \, y) \log\frac{\Phi(x \, | \, y)}{{U\mathstrut}^{*}(x)}
\\ + \\
\sum_{x, \, y}{U\mathstrut}^{*}(x, \, y) \log\frac{{U\mathstrut}^{*}(x \, | \, y)}{\Phi(x \, | \, y)}}}
\;\;\,
D\big({U\mathstrut}_{x}^{*} \circ {U\!\mathstrut}_{y \, | \, x}\,\|\,{U\mathstrut}_{x}^{*}\circ P\big)
\nonumber
\end{align}
\begin{align}
& \!\overset{(b)}{\geq}
\;\;\;
\min_{\substack{\\U(y \, | \, x):\\ \sum_{x, \, y}{U\mathstrut}^{*}(x)U(y \, | \, x) \log\frac{{U\mathstrut}^{*}(x \, | \, y)}{{U\mathstrut}^{*}(x)} \\ \geq \\ \sum_{x, \, y}{U\mathstrut}^{*}(x, \, y) \log\frac{\Phi(x \, | \, y)}{{U\mathstrut}^{*}(x)}}}
\;\;\,
D\big({U\mathstrut}_{x}^{*} \circ {U\!\mathstrut}_{y \, | \, x}\,\|\,{U\mathstrut}_{x}^{*}\circ P\big)
\nonumber \\
& \!\overset{(c)}{\geq}
\min_{\substack{\\U(y \, | \, x):\\ \sum_{x, \, y}{U\mathstrut}^{*}(x)U(y \, | \, x) \log\frac{{U\mathstrut}^{*}(x \, | \, y)}{{U\mathstrut}^{*}(x)} \; \geq \; T}}
\!
D\big({U\mathstrut}_{x}^{*} \circ {U\!\mathstrut}_{y \, | \, x}\,\|\,{U\mathstrut}_{x}^{*}\circ P\big)
\nonumber \\
& \!\overset{(d)}{\geq}
\!\!\min_{\substack{\\U(x, \, y):\\ \sum_{x, \, y}U(x, \, y) \log\frac{{U\mathstrut}^{*}(x \, | \, y)}{U(x)} \; \geq \; T}}
\!\!\!\!\!\!\!
\Big\{
D({U\!\mathstrut}_{x,\,y} \,\|\, {U\mathstrut}^{*}_{x}\circ P)\Big\}
\nonumber \\
& = \;\; \hat{E}\big(T, \,{U\mathstrut}^{*}_{x}, \,{U\mathstrut}^{*}_{x \, | \, y}\big),
\nonumber
\end{align}
where ${U\mathstrut}^{*}(x \, | \, y)$ can be considered as a stochastic matrix defined only for $\{y:\,{U\mathstrut}^{*}(y) > 0\}$,
or, alternatively, as defined arbitrarily
or extended with $\Phi(x \, | \, y)$ for $\{y:\,{U\mathstrut}^{*}(y) = 0\}$;\newline
(a) holds because ${U\mathstrut}^{*}(y\,|\,x)$ satisfies the condition under $\min$;\newline
(b) holds because $\sum_{x, \, y}{U\mathstrut}^{*}(x, \, y) \log\frac{{U\mathstrut}^{*}(x \, | \, y)}{\Phi(x \, | \, y)}\,\geq\, 0$;\newline
(c) holds because $\sum_{x, \, y}{U\mathstrut}^{*}(x, \, y) \log\frac{\Phi(x \, | \, y)}{{U\mathstrut}^{*}(x)}\,\geq\,T$;\newline
(d) holds because of the further minimization over $U(x)$.
\end{proof}

\bigskip

We conclude from Lemma~\ref{lemma1}, that, given $\hat{E}(T, {Q\mathstrut}_{0}, {\Phi\mathstrut}_{0}) < \infty$, the sequence $\big\{\hat{E}(T, {Q\mathstrut}_{l}, {\Phi\mathstrut}_{l})\big\}_{l\, = \, 0}^{+\infty}\;$
converges.
Now, it is desirable to know -- when this sequence converges all the way to zero, and when it is stuck at some positive level.
We distinguish between two cases by comparing
the threshold $T$ to the channel capacity $C$.

\bigskip

{\em Proposition 2:}
{\em If $\,T > C$, then the sequence \newline $\big\{\hat{E}(T, {Q\mathstrut}_{l}, {\Phi\mathstrut}_{l})\big\}_{l\, = \, 0}^{+\infty}\;$ cannot decrease to zero.}

\bigskip

\begin{proof}
Observe from (\ref{eqErasureExponent}), that $\hat{E}(T, Q, \Phi)$ is a non-decreasing function of $T$, 
positive
for $T$ greater than
\begin{align}
{T\mathstrut}_{0}\, & = \,\sum_{x, \, y}Q(x)P(y\,|\,x) \log\frac{\Phi(x\,|\,y)}{Q(x)}
\label{eqT0} \\
& \leq \,
\sum_{x, \, y}Q(x)P(y\,|\,x) \log\frac{(Q\circ P)(x\,|\,y)}{Q(x)}
= I(Q\circ P) \leq C.
\nonumber
\end{align}
\end{proof}

For the case $T \,\leq \,C$ we need two lemmas first.
\bigskip
\begin{lemma} \label{lemma2}
{\em Let $\big\{({Q\mathstrut}_{l_{i}}\, , \, {\Phi\mathstrut}_{l_{i}} )\big\}_{i \, = \, 1}^{+\infty}$ be a converging subsequence:}
\begin{displaymath}
{Q\mathstrut}_{l_{i}}(x) \; \overset{i \, \rightarrow \, \infty}{\longrightarrow} \; {\overline{\!Q\mathstrut}}(x),
\;\;\;\;\;
{\Phi\mathstrut}_{l_{i}}(x \, | \, y) \; \overset{i \, \rightarrow \, \infty}{\longrightarrow} \; {\overline{\Phi\mathstrut}}(x \, | \, y).
\end{displaymath}
{\em Then also}
\begin{equation} \label{eqSubseq}
{Q\mathstrut}_{l_{i}\, + \, 1}(x) \; \overset{i \, \rightarrow \, \infty}{\longrightarrow} \; {\overline{\!Q\mathstrut}}(x),
\;\;\;\;\;
{\Phi\mathstrut}_{l_{i}\, + \, 1}(x \, | \, y) \; \overset{i \, \rightarrow \, \infty}{\longrightarrow} \; {\overline{\Phi\mathstrut}}(x \, | \, y).
\end{equation}
\end{lemma}

\bigskip

\begin{proof}
From the proof of the previous lemma, it is clear that at each iteration the minimum $\hat{E}(T, {Q\mathstrut}_{l}, {\Phi\mathstrut}_{l})$
decreases by at least the amount $D({Q\mathstrut}_{l\, + \, 1} \, \| \, {Q\mathstrut}_{l})$:
\begin{displaymath}
\hat{E}(T, {Q\mathstrut}_{l}, {\Phi\mathstrut}_{l}) - \hat{E}(T, {Q\mathstrut}_{l\, + \,1}, {\Phi\mathstrut}_{l\,+\,1})
\geq D({Q\mathstrut}_{l\, + \, 1} \, \| \, {Q\mathstrut}_{l})  \overset{l \, \rightarrow \, \infty}{\longrightarrow} 0.
\end{displaymath}
Therefore ${Q\mathstrut}_{l_{i}\,+\,1}$ converges to the same limit as ${Q\mathstrut}_{l_{i}}$.

Similarly for ${\Phi\mathstrut}_{l\, + \, 1}\,$. Observe that if strict inequality holds:
\begin{displaymath}
\sum_{x, \, y} {U\mathstrut}^{*}_{l}(x, y) \log \frac{{U\mathstrut}^{*}_{l}(x\,|\,y)}{{U\mathstrut}^{*}_{l}(x)}
>
\sum_{x, \, y} {U\mathstrut}^{*}_{l}(x, y) \log \frac{{\Phi\mathstrut}_{l}(x\,|\,y)}{{U\mathstrut}^{*}_{l}(x)}
\geq T,
\end{displaymath}
then, in case the divergence in the minimum (\ref{eqErasureExponent}), with ${\Phi\mathstrut}_{l}$, is positive for ${U\mathstrut}^{*}_{l}(x, y)$, it can be further decreased with the choice ${\Phi\mathstrut}_{l\,+\,1}(x\,|\,y) = {U\mathstrut}^{*}_{l}(x\,|\,y) \neq {\Phi\mathstrut}_{l}(x\,|\,y)$.
In case the minimum (\ref{eqErasureExponent}) with ${\Phi\mathstrut}_{l}$ is exactly zero, ${\Phi\mathstrut}_{l}$ becomes constant after a single update (\ref{eqUpdatePhi}). We conclude, that in any case ${\Phi\mathstrut}_{l_{i}\, + \, 1}$ has to converge to the same limit as ${\Phi\mathstrut}_{l_{i}}$.
\end{proof}

Finally, we need the explicit solution of (\ref{eqErasureExponent}), given by

\bigskip

\begin{lemma} \label{lemma3}
\begin{equation} \label{eqExplicit}
\hat{E}(T, Q, \Phi) \;
\equiv \;
\sup_{\rho \, > \, 0} \big\{{\hat{E}\mathstrut}_{0}(\rho, Q, \Phi) + \rho T\big\},
\end{equation}
{\em where}
\begin{align}
& {\hat{E}\mathstrut}_{0}(\rho, Q, \Phi) \; \triangleq
\nonumber \\
&  - (1+\rho)\,\log\, \sum_{x}\bigg[Q(x)\sum_{y}P(y\,|\,x)\Phi^{\rho}(x\,|\,y)\bigg]^{\frac{1}{1\,+\,\rho}},
\label{eqE0}
\end{align}
{\em and if the minimum is finite, then the minimizing distribution is given by}
\begin{align}
{U\!\!\mathstrut}_{\rho}(x) \; & \propto \; \bigg[Q(x)\sum_{y}P(y\,|\,x){\Phi\mathstrut}^{\rho}(x\,|\,y)\bigg]^{\frac{1}{1\,+\,\rho}},
\label{eqUrho} \\
{U\!\!\mathstrut}_{\rho}(y \,|\,x) \; & \propto \; P(y\,|\,x){\Phi\mathstrut}^{\rho}(x\,|\,y),
\label{eqUCondrho}
\end{align}
{\em for some $\rho \in [0, +\infty]$.}
\end{lemma}

\bigskip

\begin{proof}
\begin{align}
& \;\;\;\;\;\;\;\,
\min_{\substack{\\U(x, \, y):\\\sum_{x, \, y}U(x, \, y) \log\frac{\Phi(x \, | \, y)}{U(x)} \; \geq \; T}}
\,\Big\{D({U\!\mathstrut}_{x, \, y} \,\|\,Q\circ P)\Big\}
\nonumber \\
&
\overset{\rho \, > \, 0}{\geq}
\min_{\substack{\\U(x, \, y):\\\sum_{x, \, y}U(x, \, y) \log\frac{\Phi(x \, | \, y)}{U(x)} \; \geq \; T}}
\Bigg\{
D({U\!\mathstrut}_{x, \, y} \,\|\,Q\circ P)
\nonumber \\
&
\;\;\;\;\;\;\;\;\;\;\;\;\;\;\;\;\;\;\;\;\;\;\;\;\;\;\;\;\;\;\;\,
- \,
\rho \underbrace{\Bigg[\sum_{x, \, y}U(x, y) \log \frac{\Phi(x\,|\,y)}{U(x)}\,-\,T\Bigg]}_{\geq \, 0}
\Bigg\}
\nonumber \\
& \;\, \geq
\;\;\;\;\;\;\;\;\;\;\;\;\;\;\;\,
\min_{\substack{\\U(x, \, y)}}
\;\;\;\;\;\;\;\;\;\;\;\;\;\,
\Bigg\{
D({U\!\mathstrut}_{x, \, y} \,\|\,Q\circ P)
\nonumber \\
&
\;\;\;\;\;\;\;\;\;\;\;\;\;\;\;\;\;\;\;\;\;\;\;\;\;\;\;\;\;\;\;\,
- \,
\rho\,\Bigg[\sum_{x, \, y}U(x, y) \log \frac{\Phi(x\,|\,y)}{U(x)} \,-\, T\Bigg]\,
\Bigg\}
\nonumber \\
& \;\, =
\;\;\;\;\;\;\;\;\;\;\;\;\;\;\;\,
\min_{\substack{\\U(x, \, y)}}
\;\;\;\;\;\;\;\;\;\;\;\;\;\,
\Bigg\{
(1+\rho)\underbrace{\sum_{x}U(x)\log\frac{U(x)}{{U\!\!\mathstrut}_{\rho}(x)}}_{\geq\,0}
\nonumber \\
&
\;\;
+ {\hat{E}\mathstrut}_{0}(\rho, Q, \Phi) +
\sum_{x}U(x)\underbrace{\sum_{y} U(y\,|\,x) \log \frac{U(y\,|\,x)}{{U\!\!\mathstrut}_{\rho}(y\,|\,x)}}_{\geq\,0} + \,\rho T
\Bigg\}
\nonumber \\
& \;\, = \;
{\hat{E}\mathstrut}_{0}(\rho, Q, \Phi)\, + \, \rho T
\nonumber
\end{align}
\begin{align}
& \;\, \geq \;\;\;\;\,
\min_{\substack{\\U(x, \, y):\\\sum_{x, \, y}U(x, \, y) \log\frac{\Phi(x \, | \, y)}{U(x)} \\ \geq \\
\sum_{x, \, y}{U\!\!\mathstrut}_{\rho}(x, \, y) \log\frac{\Phi(x \, | \, y)}{{U\!\!\mathstrut}_{\rho}(x)}}}
\;\;\;
\Big\{
D({U\!\mathstrut}_{x, \, y} \,\|\,Q\circ P)\Big\}
\nonumber \\
&
\;\;\;\;\;\;\;\;\;\;\;\;\;\;\;\;\;\;\;\;\;\;\;\;\;\;\;\;\;\;\;\,
- \,
\rho\,\Bigg[\sum_{x, \, y}{U\!\!\mathstrut}_{\rho}(x, y) \log \frac{\Phi(x\,|\,y)}{{U\!\!\mathstrut}_{\rho}(x)}\,-\,T\Bigg].
\nonumber
\end{align}
From the string of inequalities above, we see that the LHS of (\ref{eqExplicit}), $\hat{E}(T, Q, \Phi)$, as a function of $T$, is lower-bounded by the straight lines $\,E(T) \, = \, {\hat{E}\mathstrut}_{0}(\rho, Q, \Phi) + \rho T\,$
of slopes $\rho > 0$. The last inequality shows in particular, that for each $\rho$ there exists $T$ -- where the line $\, {\hat{E}\mathstrut}_{0}(\rho, Q, \Phi) + \rho T\,$
touches the curve $\hat{E}(T, Q, \Phi)$. We conclude that the supremum on the RHS of (\ref{eqExplicit}) is the {\em lower convex envelope} of $\hat{E}(T, Q, \Phi)$.

On the other hand, it can be checked directly by the definition of convexity, that the LHS of (\ref{eqExplicit}) is a convex ($\cup$) function of $T$.
Therefore, the LHS of (\ref{eqExplicit}) must coincide with its lower convex envelope, which is given by the RHS.
\end{proof}

\bigskip

Our main result is given by the following.

\bigskip

\begin{thm} \label{thm1}
{\em Let $\big\{{U\mathstrut}^{*}_{l}\big\}_{l\,=\,0}^{+\infty}\,$ be a sequence of iterative solutions of (\ref{eqErasureExponent}),
given by (\ref{eqUrho}, \ref{eqUCondrho}) with $Q = {Q}_{l}\,$, $\Phi = {\Phi}_{l}\,$, and $\rho = {\rho\mathstrut}_{l}\,$,
such that the following conditions hold:}
\begin{align}
\limsup_{l\,\rightarrow\,\infty}\;{\rho\mathstrut}_{l} \; & < \; 1,
\label{eqRhounderone} \\
\liminf_{l\,\rightarrow\,\infty}\;{Q}_{l}(x) \; & > \; 0, \;\;\;\;\;\; \forall \; x \, \in \, {\cal X},
\label{eqQbounded} \\
\liminf_{l\,\rightarrow\,\infty}\;{\Phi}_{l}(x \, | \, y) \; & > \; 0, \;\;\;\;\;\; \forall \; (x, y): \; P(y \, | \, x) \, > \, 0.
\label{eqPhibounded}
\end{align}
{\em Then the sequence $\big\{\hat{E}(T, {Q\mathstrut}_{l}, {\Phi\mathstrut}_{l})\big\}_{l\, = \, 0}^{+\infty}\;$ converges to zero if $\,T < C$.}
\end{thm}

\bigskip

\begin{proof}
Suppose $\,T < C\,$ and $\,\big\{\big({Q\mathstrut}_{l_{i}}\, , \, {\Phi\mathstrut}_{l_{i}}\, , \, {\rho\mathstrut}_{l_{i}} \big)\big\}_{i \, = \, 1}^{+\infty}\,$ is a converging subsequence:
\begin{displaymath}
{Q\mathstrut}_{l_{i}}(x) \overset{i \, \rightarrow \, \infty}{\longrightarrow} \,{\overline{\!Q\mathstrut}}(x),
\;\;\;
{\Phi\mathstrut}_{l_{i}}(x \, | \, y) \overset{i \, \rightarrow \, \infty}{\longrightarrow} \,{\overline{\Phi\mathstrut}}(x \, | \, y),
\;\;\;
{\rho\mathstrut}_{l_{i}} \overset{i \, \rightarrow \, \infty}{\longrightarrow} \bar{\rho},
\end{displaymath}
such that the limit of ${\rho\mathstrut}_{l_{i}}$ is positive $0 \, < \, \bar{\rho} \, < \, 1$.
Then, by Lemma~\ref{lemma2}, continuity of (\ref{eqUrho}), and using boundedness (\ref{eqQbounded}), we obtain:
\begin{align}
{\overline{\!Q\mathstrut}}(x) \; & \overset{(\ref{eqSubseq})}{=} \; \lim_{i\,\rightarrow\,\infty}{Q\mathstrut}_{l_{i}\, + \, 1}(x)
\; \overset{(\ref{eqUpdateQ})}{\equiv} \; \lim_{i\,\rightarrow\,\infty}{U\mathstrut}_{l_{i}}^{*}(x)
\label{eqQLimit} \\
& \!\overset{(\ref{eqUrho})}{\propto} \;
\bigg[\,{\overline{\!Q\mathstrut}}(x)\sum_{y}P(y\,|\,x)\,{{\overline{\Phi\mathstrut}}}^{\,\bar{\rho}}(x\,|\,y)\bigg]^{\frac{1}{1\,+\,\bar{\rho}}}
\; \overset{(\ref{eqQbounded})}{>} \; 0,
\nonumber \\
{\overline{\!Q\mathstrut}}(x) \; & \,\propto \;
\bigg[\sum_{y}P(y\,|\,x)\,{{\overline{\Phi\mathstrut}}}^{\,\bar{\rho}}(x\,|\,y)\bigg]^{1/\bar{\rho}}.
\label{eqQArimoto}
\end{align}
Also for the limit ${\overline{\Phi\mathstrut}}(x \, | \, y)$, with the help of (\ref{eqQArimoto}), by continuity of (\ref{eqUCondrho}),
using Lemma~\ref{lemma2} and condition (\ref{eqPhibounded}):
\begin{align}
\lim_{i\,\rightarrow \, \infty}{U\mathstrut}^{*}_{l_{i}}(x, y) \;\; \!\overset{(\ref{eqQLimit}, \, \ref{eqUCondrho})}{=}
& \;\;
\,{\overline{\!Q\mathstrut}}(x)\,
\frac{P(y\,|\,x)\,{\overline{\Phi\mathstrut}}^{\,\bar\rho}(x\,|\,y)}{\sum_{b}P(b\,|\,x)\,{{\overline{\Phi\mathstrut}}}^{\,\bar{\rho}}(x\,|\,b)}
\nonumber \\
\overset{(\ref{eqQArimoto})}{\propto} \;\;
& \;\;
\,{\overline{\!Q\mathstrut}}^{\,1\,-\,\bar{\rho}}(x)\,
P(y\,|\,x)\,{\overline{\Phi\mathstrut}}^{\,\bar\rho}(x\,|\,y)
\label{eqJointLimit} \\
{\overline{\Phi\mathstrut}}(x \, | \, y)
\;\; \overset{(\ref{eqSubseq})}{=} \;\;\,
&
\lim_{i\,\rightarrow\,\infty}{\Phi\mathstrut}_{l_{i}\,+\,1}(x \, | \, y)
\nonumber
\end{align}
\begin{align}
\overset{(\ref{eqUpdatePhi}, \, \ref{eqPhibounded})}{=} \,
& \lim_{i\,\rightarrow\,\infty}{U\mathstrut}^{*}_{l_{i}}(x \, | \, y)
\nonumber \\
\overset{(\ref{eqJointLimit})}{\propto} \;\; &
\;\;
\,{\overline{\!Q\mathstrut}}^{\,1\,-\,\bar{\rho}}(x)\,
P(y\,|\,x)\,{\overline{\Phi\mathstrut}}^{\,\bar\rho}(x\,|\,y)
\nonumber \\
{\overline{\Phi\mathstrut}}(x \, | \, y)
\; \overset{(\ref{eqPhibounded})}{\propto} \;\; & \;\;\,{\overline{\!Q\mathstrut}}(x)
P^{\frac{1}{1\,-\,\bar{\rho}}}(y\,|\,x).
\label{eqPhiArimoto}
\end{align}
The expressions (\ref{eqQArimoto}) and (\ref{eqPhiArimoto}) can be recognized as Arimoto's minimization solutions \cite[eq.~(11),~(9)]{Arimoto76},
both satisfied at the same time, implying the minimization of the exponent function ${E\mathstrut}_{0}(-\bar{\rho}, Q)$ by $\,{\overline{\!Q\mathstrut}}$.
Equivalently, if we plug ${\overline{\Phi\mathstrut}}(x \, | \, y)$, as given by (\ref{eqPhiArimoto}), into the expression of $\,{\overline{\!Q\mathstrut}}(x)$ (\ref{eqQArimoto}),
and reduce $\,{\overline{\!Q\mathstrut}}(x)$ on both sides,
we arrive at the condition
\begin{equation} \label{eqSufficientCondition}
\sum_{y}P^{\frac{1}{1\,-\,\bar{\rho}}}(y\,|\,x) \bigg[\sum_{a}\, {\overline{\!Q\mathstrut}}(a)
P^{\frac{1}{1\,-\,\bar{\rho}}}(y\,|\,a)\bigg]^{-\bar{\rho}}
\; 
=
\; \underset{\substack{\\ \\\forall \; x \, \in \, {\cal X}}}{const}.
\end{equation}
This is a sufficient condition for $\,{\overline{\!Q\mathstrut}}$ to minimize ${E\mathstrut}_{0}(-\bar{\rho}, Q)$ \cite[eq.~22]{Arimoto76}.
Using (\ref{eqSufficientCondition}) in the definition (\ref{eqE0}) gives
\begin{displaymath}
{\hat{E}\mathstrut}_{0}\big(\bar{\rho}, \,{\overline{\!Q\mathstrut}}, {\overline{\Phi\mathstrut}}\big) \; = \; {E\mathstrut}_{0}(-\bar{\rho}, \,{\overline{\!Q\mathstrut}}).
\end{displaymath}
On the other hand, by continuity of (\ref{eqE0}) we have
\begin{displaymath}
\lim_{i\,\rightarrow\,\infty}
\left\{{\hat{E}\mathstrut}_{0}\big({\rho\mathstrut}_{l_{i}}, {Q\mathstrut}_{l_{i}}, {\Phi\mathstrut}_{l_{i}}\big) \, + \, {\rho\mathstrut}_{l_{i}} T\right\}
\; = \;
{\hat{E}\mathstrut}_{0}\big(\bar{\rho}, \,{\overline{\!Q\mathstrut}}, {\overline{\Phi\mathstrut}}\big) \, + \, \bar{\rho} T,
\end{displaymath}
which is also the limit of the monotonically non-increasing sequence $\big\{\hat{E}(T, {Q\mathstrut}_{l}, {\Phi\mathstrut}_{l})\big\}_{l\, = \, 0}^{+\infty}\,$, therefore it follows that
\begin{equation} \label{eqTouch}
\lim_{l\,\rightarrow\,\infty}
\hat{E}(T, {Q\mathstrut}_{l}, {\Phi\mathstrut}_{l}) \; = \; \min_{Q} {E\mathstrut}_{0}(-\bar{\rho}, Q) \, + \, \bar{\rho} T
\; \geq \; 0.
\end{equation}
Observe, however, that the straight line of the positive slope $\bar{\rho}$ on the RHS of (\ref{eqTouch}) cannot cross the $T$-axis below the capacity $C$. So it must cross the $T$-axis above the capacity, which is in {\em contradiction} to the condition $T \, < \, C$.

Therefore, given (\ref{eqRhounderone}), there {\em does not exist} a subsequence $\big\{{\rho\mathstrut}_{l_{i}}\big\}_{i\,=\,0}^{+\infty}$,
converging to a positive value $\bar{\rho}$.
We conclude that $\,\lim_{l\,\rightarrow\,\infty}{\rho\mathstrut}_{l}\, = \, 0\,$ and by (\ref{eqUrho})-(\ref{eqUCondrho})
\begin{displaymath}
D\big({U\mathstrut}^{*}_{l}(x) \, \| \, {Q\mathstrut}_{l}(x)\big) \; \overset{l\,\rightarrow\,\infty}{\longrightarrow} \; 0,
\;\;\;\;\;\;\;\;\;
{U\mathstrut}^{*}_{l}(y\,|\,x) \; \overset{l\,\rightarrow\,\infty}{\longrightarrow} \; P(y \, | \, x),
\end{displaymath}
implying $\hat{E}(T, {Q\mathstrut}_{l}, {\Phi\mathstrut}_{l})\,\searrow\, 0$.
\end{proof}

\begin{figure}[t]
\centering
\includegraphics[width=3.49in]{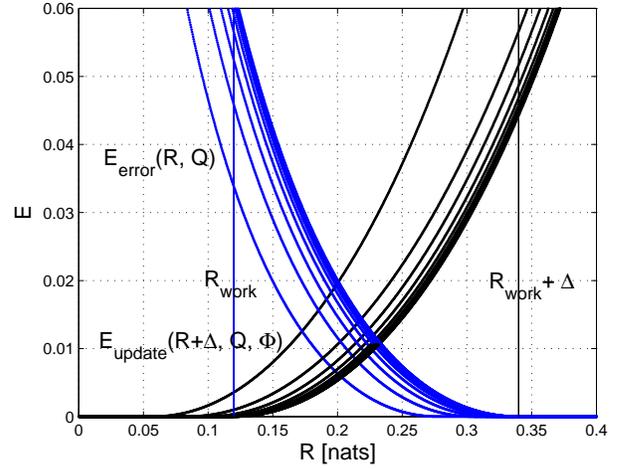}  
\caption{Example of iterations.}
\label{fig2}
\end{figure}

An example of convergence is shown in Fig.~\ref{fig2}. The rate of communication is ${R\mathstrut}_{\text{work}}$.
The threshold is $T = {R\mathstrut}_{\text{work}} + \Delta$.
The error exponent ${E}_{r}({R\mathstrut}_{\text{work}}, Q)$ is
well above the update exponent $\hat{E}({R\mathstrut}_{\text{work}}+\Delta, \,Q, \,\Phi)$,
which converges to zero. As the update exponent converges,
the zero point of the error exponent at $R = I(Q\circ P)$ moves towards $R = {R\mathstrut}_{\text{work}}+\Delta$.

\bigskip

\section{Discussion} \label{Discussion}
\bigskip

Observe, that if the update exponent (\ref{eqErasureExponent}) is zero at $T$,
then necessarily $T \leq {T\mathstrut}_{0} \leq I(Q\circ P)$ by (\ref{eqT0}), and the error exponent
(\ref{eqErrorExponent}) is positive for $R < T$.
This makes reliable communication possible at $R$, which is our goal.

Suppose, at some initial point in time, the system is in a reliable communication mode, with a rate $R<I(Q\circ P)$, and we choose the threshold $T$, $R < T \leq I(Q\circ P)$, and the stochastic matrix $\Phi$, such that the update exponent (\ref{eqErasureExponent}) is zero.
This is possible by choosing initially, for example, $\Phi(x\,|\,y) = (Q\circ P)(x\,|\,y)$.
Then, a small 
change in the channel $P(y\,|\,x)$ occurs,
so that the update exponent rises slightly above zero,
but is still lower than the decoding error exponent.
Our basic assumption is that the last condition will remain satisfied
after each subsequent iteration of the algorithm described in Section~\ref{Scheme},
so that by Proposition~1, with high probability the
update will continue according to the optimal solution of (\ref{eqErasureExponent}), as in (\ref{eqUpdateQ})-(\ref{eqUpdatePhi}).

Since the update exponent is relatively low and the change in the channel is small,
we make
another assumption -- that the repeated iteration of the updates (\ref{eqUpdateQ})-(\ref{eqUpdatePhi})
will produce a sequence $\big\{({Q\mathstrut}_{l}, {\Phi\mathstrut}_{l})\big\}$, satisfying the conditions of Theorem~\ref{thm1}.
Specifically, the slope $\rho$ of the update exponent at $T$ will remain small and the sequence $\big\{({Q\mathstrut}_{l}, {\Phi\mathstrut}_{l})\big\}$
will 
not stray to zero on some $x\in {\cal X}$.
Finally, if the new capacity $C$ is still higher than $T$, the iterations will converge by Theorem~\ref{thm1}.
The system will return to the initial state with zero update exponent (\ref{eqErasureExponent}), with respect to the new channel $P(y\,|\,x)$.

In this way, the adaptation scheme will safeguard the reliable communication mode for as long as the channel capacity $C$ doesn't go below $T$.


\bibliographystyle{IEEEtran}

\end{document}